\documentclass[twocolumn,showpacs,preprintnumbers,amsmath,amssymb]{revtex4}
\usepackage{graphicx}% Include figure files
\usepackage{dcolumn}% Align table columns on decimal point
\usepackage{bm}% bold math
 
\def\pbn{$\overline{p}$}
\def\nu{$\overline{N}\;$}

\def\pb{$\overline{p}\;$}
\def\nb{$\overline{n}\;$}

\begin{document}
 
\preprint{Report ISN 02.80}

\input epsf
\title{Secondary antiproton flux induced by cosmic ray interactions with the atmosphere}

\author{C.Y. Huang \footnote{Present address: MPIK, Saupfercheckweg 1, D-69117 Heidelberg, 
Germany}, L. Derome, and M. Bu\'enerd \footnote{Corresponding author: 
buenerd@in2p3.fr}} 
\affiliation{Laboratoire de Physique Subatomique et de Cosmologie, IN2P3/CNRS, 53 Av. des 
Martyrs, 38026 Grenoble-cedex, France}

\date{\today}% It is always \today, today,
             %  but any date may be explicitly specified\begin{flushright}

\begin{abstract}
The atmospheric secondary antiproton flux is studied for detection altitudes extending from 
sea level up to about 3 earth radii, by means of a 3-dimensional Monte-Carlo simulation, 
successfully applied previously on other satellite and balloon data. 
The calculated antiproton flux at mountain altitude is found in fair agreement with the 
recent BESS measurements. The flux obtained at balloon altitude is also in agreement with 
calculations performed in previous studies and used for the analysis of 
balloon data. The flux at sea level is found to be significant. The antineutron flux is 
also evaluated. The antiproton flux is prospectively explored up to around 2 10$^4$~km. 
The results are discussed in the context of the forthcoming measurements by large acceptance 
experiments.

\end{abstract}

%94.30.Hn Trapped particles 
%95.85.Ry Neutrino, muon, pion, and other elementary particles; cosmic rays 
%96.40.-z Cosmic rays
%13.85.-t Hadron-induced high- and super-high-energy interactions (energy>10 GeV) 
\maketitle 
\section{Introduction}
The antiproton (\pbn) has a particular status in the spectrum of Cosmic Radiation mainly 
because of 
its particular production dynamics and kinematics. The main part of the Cosmic Ray (CR) \pb 
spectrum measured in balloon and satellite experiments is well accounted for by assuming it
to consist of a secondary flux, originating from the interaction between the nuclear CR flux 
and the interstellar matter in the galaxy (ISM) \cite{DO01}. It is expected to be dominant 
with respect to components from other possible origins. Such other contributions of primary 
origin and of major astrophysical interest, have been considered recently. In particular, the 
\pb flux induced by annihilation of dark matter constituents \cite{ST88,BO99a,BE99}, and by 
primordial black hole evaporation \cite{CA76,BA00} have been discussed. All these possible 
contributions are intimately entangled together and their phenomenological disentagling 
relies critically on the accuracy of the experimental data. 
The measurements of the \pb flux thus provide a sensitive test of the production source and 
mechanism, and of the propagation conditions in the galaxy 
\cite{DO01,GA92,CH97,SI98,BI99,UL99,MO02}.
\par
CR antiprotons have been experimentally studied for several decades by satellite or balloon 
borne experiments (see references in \cite{BO01}). Several recent balloon  experiments, 
like BESS \cite{BESS97,BESS98} and CAPRICE \cite{BO01,BO97} have collected new data samples 
whose analysis have provided determinations of the galactic \pb flux over a kinetic energy 
range extending from about 0.2~GeV kinetic energy up to about 50~GeV. In these works, the 
values of the antiproton galactic flux were obtained by subtracting the calculated 
atmospheric \pb flux from the values of the measured total flux. 
\par
Secondary galactic as well as atmospheric antiprotons are both produced in hadronic 
collisions by the same elementary reaction mechanism in nucleon-nucleon collisions 
between the incident CR flux and either ISM nuclei (mainly Hydrogen) in the galaxy, or 
atmospheric nuclei (mainly Nitrogen) in the atmosphere. The basic \pb production reaction 
is the inclusive $NN\rightarrow\bar{p}X$ process, $N$ standing for nucleon and X for any 
final quantal hadronic state allowed in the process.
The ratio of \pb production in the galaxy and in the atmosphere scales with the ratio of 
matter thickness (in units of interaction length $\lambda_I$) crossed by protons in the two 
media. These thicknesses are known to be of the same order of magnitude. In addition,
both flux are driven by similar transport equations (see \cite{GA92,SI98,PF96,GA99} for 
example), with however the escape term arising from convective and diffusive effects in 
the interstellar medium on the galactic flux \cite{GA92}, making a significant difference
with the transport of flux in atmosphere, which tends to decrease the transported flux 
compared to the atmospheric transport conditions.

It can be shown using a Leaky Box Model (LBM) for the galactic transport and a simple 
slab model for the production in the upper atmosphere \cite{MA01}, that the ratio of the 
atmospheric \pb flux at balloon altitude $N_{atm}$ to the galactic \pb flux at TOA 
$N_{gal}$, is approximately:

$$\frac{N_{atm}(\bar{p})}{N_{gal}(\bar{p})}\approx \frac{x_{Atm}}{\lambda_{e}}
\frac{\sigma(p\, Atm\rightarrow\bar{p}X)}{\sigma(pp\rightarrow\bar{p}X)}
\frac{m_H}{m_{Atm}}\approx\frac{x_{Atm}}{\lambda_{e}}(\frac{m_H}{m_{Atm}})^{1/3}$$

\noindent
Where $x_{Atm}$ is the thickness of atmosphere on top of balloon experiments, while 
$\lambda_{e}$ is the LBM escape length, $\sigma(pp\rightarrow\bar{p}X)$ and 
$\sigma(p\, Atm\rightarrow\bar{p}X)$ being the inclusive antiproton production cross 
sections on hydrogen and on the atmospheric nuclei respectively, while $m_H$ and $m_{Atm}$ 
are the hydrogen and the mean atmospheric nuclear mass respectively.
Using $x_{Atm}$=3.9~g/cm$^2$ (for a 38~km altitude) \cite{HE91} and $\lambda_{e}$=8~g/cm$^2$ 
and 11.8~g/cm$^2$ for particle rigidities of 3~GV and 10~GV respectively \cite{MATT}, the 
above ratio is found to be of the order of 0.15 and 0.2, respectively. 
The contribution of the atmospheric antiproton production to the total flux measured in 
balloon experiments is thus not expected to be negligible with respect to the galactic 
component. The correction of the total flux from the atmospheric contribution therefore 
needs the latter to be calculated very carefully since the accuracy on the evaluation of 
this component sets a limit of accuracy on the final value of the measured galactic flux. 

It must be emphasized that studying also the secondary proton flux in the atmosphere in 
this context is interesting since the latter is very sensitive to all the components of 
the simulation process, in particular to the secondary proton production cross section 
which contributes as well to the generation of the antiproton flux. The comparison of 
the calculation to the recently measured data provides a robust validation of the approach 
used and of the overall calculation, and firmly supports the reliability of the results 
reported here. This study is reported in a separate (companion) paper \cite{BA03}. 

The present work is an extension of a resarch program aiming at the interpretation of 
satellite data and which first results on the flux of protons, leptons, and light ions, 
below the geomagnetic cutoff (GC), at satellite altitude, have been reported recently 
\cite{PAP1,PAP2,PAP3}.

The paper reports on the calculated \pb atmospheric flux over the range from sea level up 
to satellite altitudes by Monte-Carlo simulation. The main features of the calculations are 
described in section \ref{SIMU}. The production cross sections used in the event generator 
are given in section \ref{CROSEC}. The results are discussed in section \ref{RESPBARS}.
The work is concluded in section \ref{CONC}. 
\section{Simulation conditions}\label{SIMU}

As mentioned above, the flux of secondary atmospheric antiprotons has been investigated 
using the same simulation approach which has allowed to successfully account for the 
$p$, $d$, $H\!e$, and $e^\pm$ experimental flux below the Geomagnetic Cutoff (GC) measured 
by the AMS experiment, as well as the experimental proton and muon flux in the atmosphere, 
the latter being studied together with the atmospheric neutrino flux \cite{LI03}. 

The same computing environment has been used here for the charged particle propagation in 
the terrestrial environment including the atmosphere, as in the previous studies, with
the event generator being dedicated however, based on the antiproton production cross 
section in nucleon-nucleon collisions. 

The calculation proceeds by means of a full 3D-simulation program. Incident Cosmic Rays are 
generated on a virtual sphere chosen at a 2000~km altitude. Random events are generated 
uniformly on this sphere. The local zenith angle distribution of the particule momentum is 
proportional to $\cos(\theta_z)d\cos(\theta_z)$, $\theta_z$ being the zenithal angle of the 
particle, in order to get an isotropic flux at any point inside the volume of the virtual 
sphere. The geomagnetic cut-off is applied by back-tracing the particle 
trajectory in the geomagnetic field, and keeping in the sample only those particles 
reaching a backtracing distance of 10 Earth radii. Flux conservation along any allowed 
particle path in the geomagnetic field is ensured by application of Liouville's theorem. 
The normal particle propagation as well as its back-tracing are performed using the 
adaptative Runge-Kutta integration method in the Geomagnetic field \cite{LI03}.

1. For the incident CR proton and helium flux, functional forms fitted to the 1998 AMS 
measurements were used \cite{PROT,HELI} (see also \cite{BO99,ME00,SA00}). The heavier 
components of the CR flux were not taken into account in the calculations (see \cite{LI03}). 
For other periods of the solar cycle, the incident cosmic flux are corrected for the 
different solar modulation effects using a simple force law approximation \cite{SOLMOD}.

2. Each particle is propagated in the geomagnetic field and interacts with nuclei of the 
local atmospheric density according to their total reaction cross section and producing 
secondary nucleons $p$, $n$, and antinucleons \pb, \nb, with cross sections and 
multiplicities. This important issue is discussed in section \ref{CROSEC} below. The specific 
ionization energy loss is computed for each step along the trajectory. 

3. In the following step, each secondary particle produced in a primary collision is 
propagated in the same conditions as incident CRs in the previous step, resulting in a more 
or less extended cascade of collisions through the atmosphere, which may include up to ten 
generations of secondaries for protons for the sample generated in this work 
\cite{BA03,PAP1}. 
%A fraction of the secondaries escaping the atmosphere could fill the belt of quasi trapped 
%particles discussed in section \ref{***}.

For the antinucleon inelastic collisions, only the annihilation reaction channel was taken 
into account. Non annihilating inelastic $\bar{N}+A\rightarrow \bar{N}+X$ ($\bar{N}$ 
standing for antinucleon) interactions whose contribution to the total reaction cross 
section $\sigma_R$, is small. It consists basically of the single diffractive dissociation cross 
section (for the proton target in individual $\bar{N}p$ collisions), and it would be of the 
same order of magnitude as for pp colisions, namely $\sim$10\% of $\sigma_R$ or less at the 
energies considered here \cite{PBARINEL}. It has been neglected at this stage. It will be 
included in the further developments of the calculation program. 
%\**cite{PBARINEL}K. Goulianos, Phys. Rep. 1001(1983)169

The reaction products are counted whenever they cross, upwards or downwards, the virtual 
detection spheres (several can be defined in the program) at the altitude of the detectors: 
from sea level up to about 36~km for ground and balloon experiments (BESS, CAPRICE), 370~km 
for the AMS satellite experiment. %Beyond for the planned satellite experiments, 
Higher altitudes up to more than 10000~km were also investigated, with the purpose of 
understanding the dynamics of the population of quasi-trapped particles in the earth 
environment (see section \ref{HIGHALT}). 
Each particle is propagated until it disappears by nuclear collision (annihilation 
for antinucleons), stopping in the atmosphere by energy loss, or escaping to the outer space 
beyond twice the generation altitude \cite{PAP1,PAP2,PAP3,LI03}.

In the terminology used in the following, one event is defined as the full cascade induced 
by an incoming CR particle interacting with one atmospheric nucleus. For each CR producing 
at least one secondary particle, the whole event is stored with all the relevant 
topological, dynamical, kinematical, and geographical informations. This includes the 
collision rank, geophysical location, altitude, momentum and particle type, and parent 
particle type, in form of event files. The collision rank is defined as the number of a given 
collision in the cascade initiated by the first CR interaction with atmosphere (rank 1). 

The calculations do not include any adjustable parameter. 

\section{Cross sections}\label{CROSEC}

\subsection{Proton induced secondaries}

\subsubsection{Protons}\label{SIGPROT}

The inclusive $p+A\rightarrow p+X$ proton production cross sections used are described in 
ref \cite{PAP1}. They are based on the results of refs \cite{BA85} and \cite{AB92} for the 
two components corresponding basically to forward (or direct quasi elastic) and backward 
(or relaxed deep inelastic) productions respectively. The values obtained have been found in 
reasonable agreement with the results of the INCL model of intranuclear cascade 
calculations \cite{BO02}. This cross section allows to reproduce very successfully the 
atmospheric secondary proton flux down to the lowest altitude \cite{BA03}.

The cross section for secondary neutron production was taken equal to that of proton production.
Similarly, the neutron induced cross sections of secondary nucleon production used was taken 
the same as for protons, the coulomb interaction making negligible differences over the 
considered energy range. 

\subsubsection{Antiprotons}

The inclusive $p+A\rightarrow\bar{p}+X$ antiproton production cross section has been 
obtained by fitting a set of available experimental data between 12 GeV incident kinetic 
energy and 24 GeV/c incident momentum \cite{SU98,AB93,AL70,EI72}, using a modified version 
of the analytical formula proposed in ref \cite{KMN}, the latter being referred to as KMN 
(for Kalinovski, Mokhov, Nikitin) in the following. 

\begin{figure*}[htbp]          %fig 1: fits pbars
\begin{center}
%\vspace*{2.0mm} 
%\includegraphics[width=8.3cm]{pbardata.eps} 
\includegraphics[width=7cm]{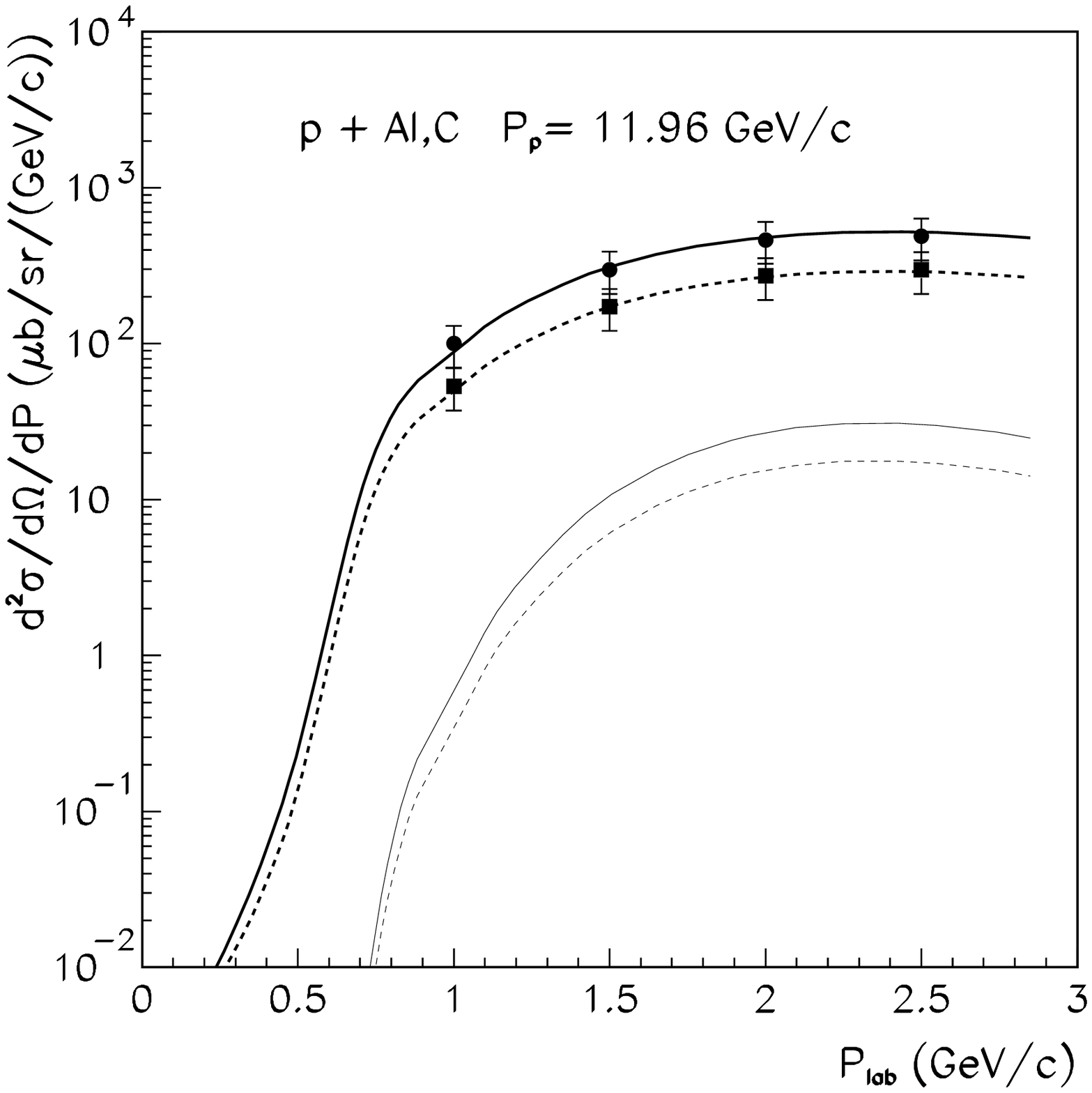} 
\includegraphics[width=7cm]{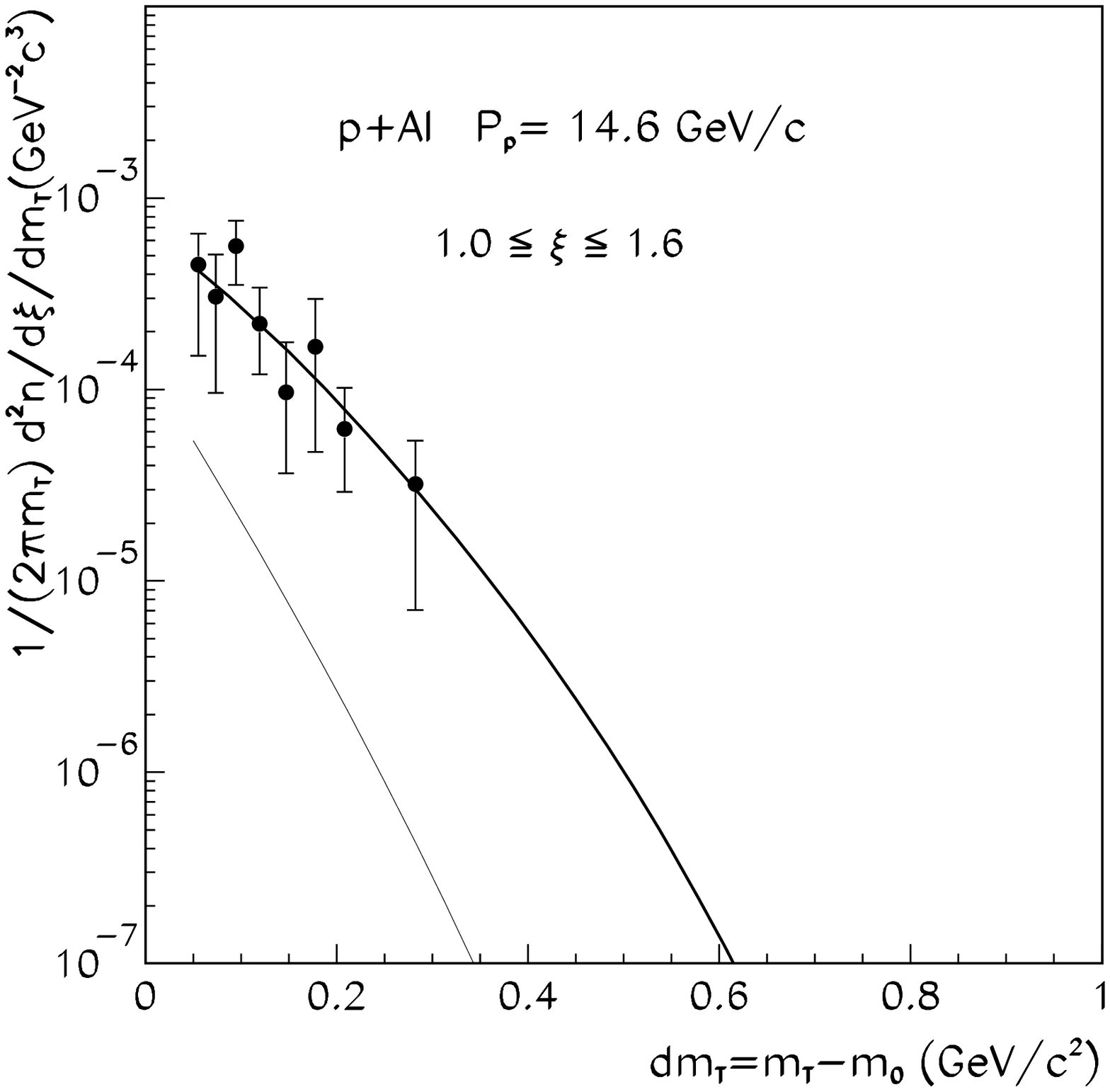} \\
\includegraphics[width=7cm]{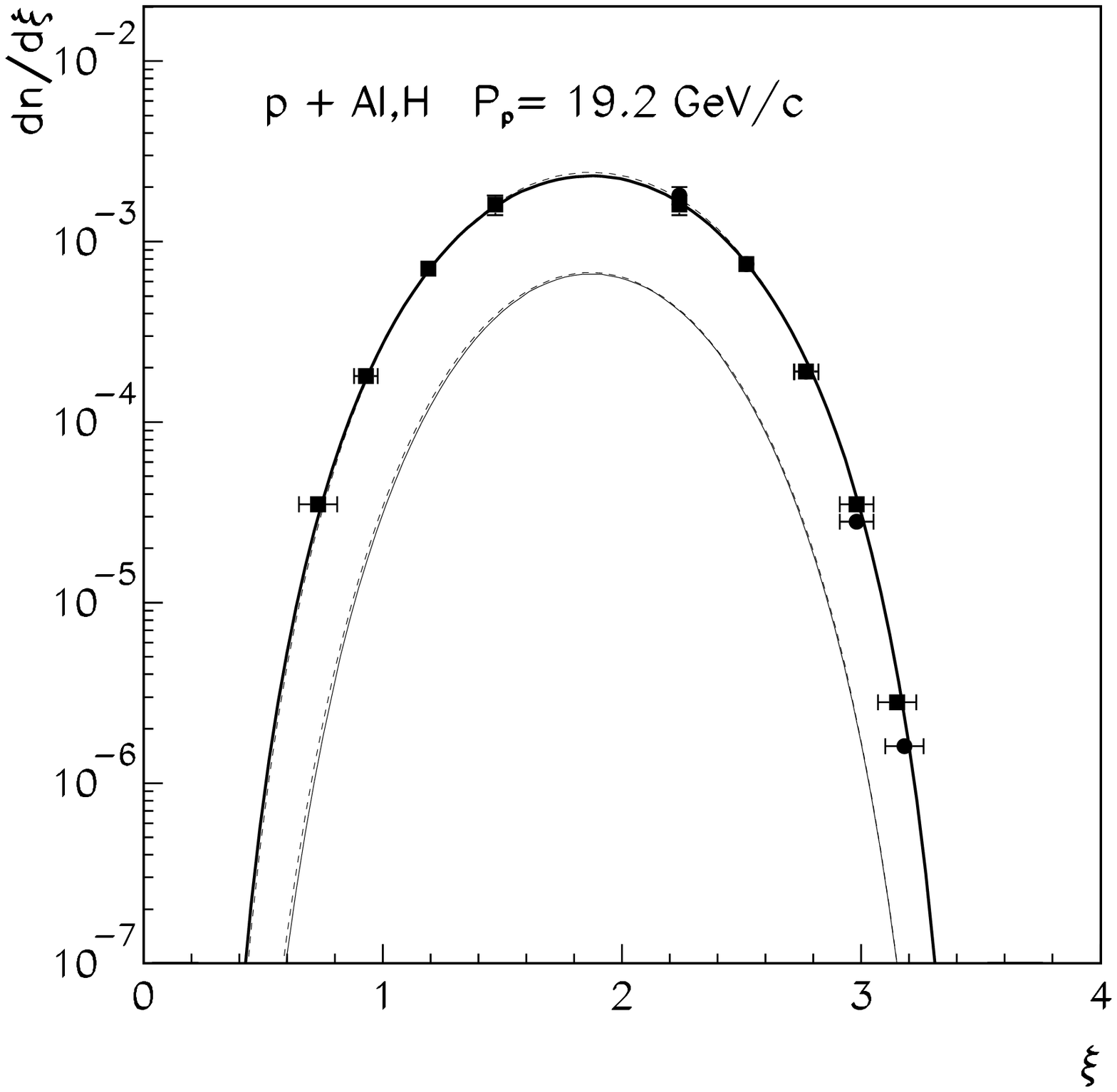} 
\includegraphics[width=7cm]{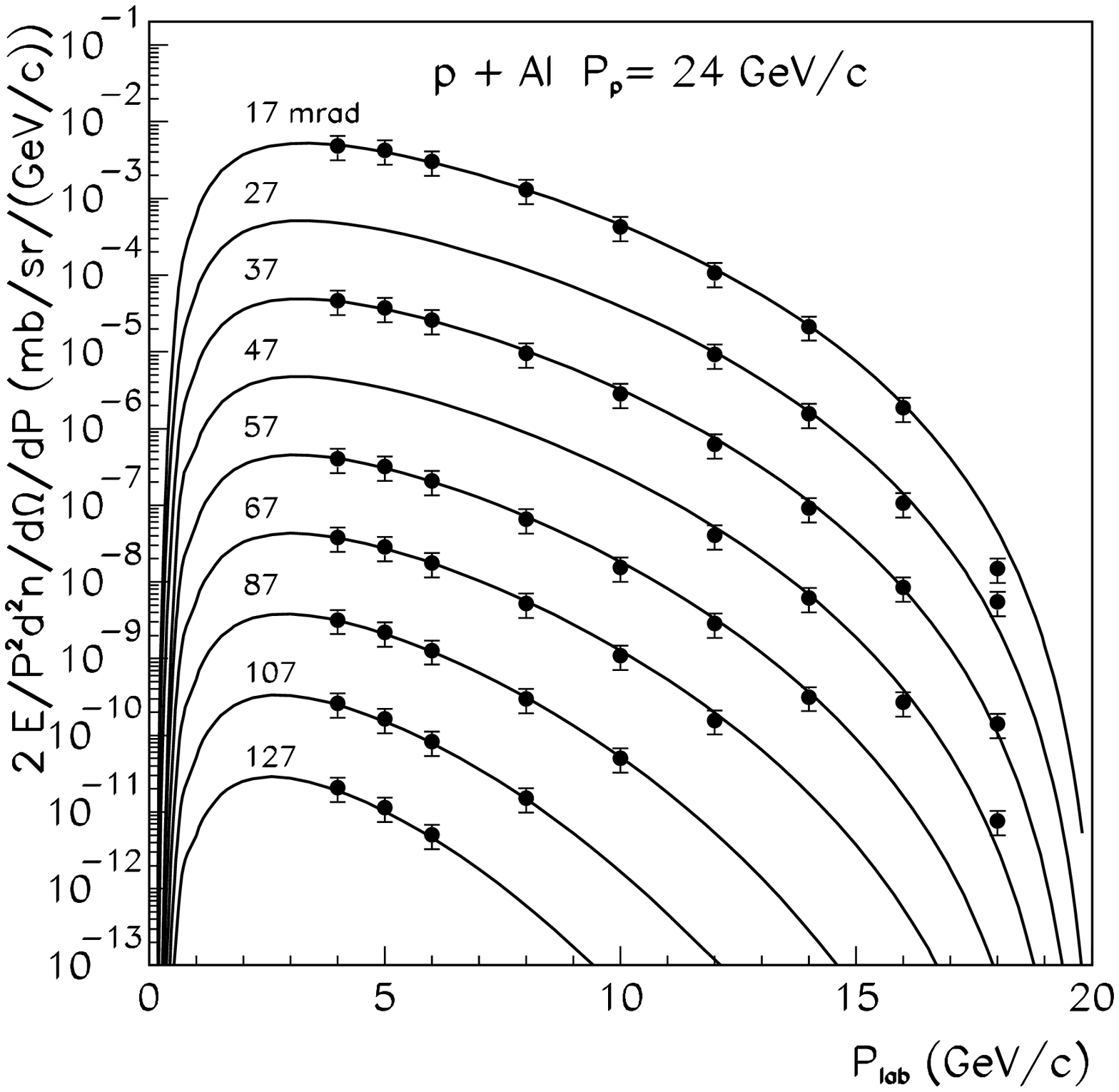} 
\caption{
\it\small Sample of the fits to the $p+A\rightarrow\bar{p}+X$ cross section data using the 
parametrized relation \ref{FKALI} given in the text, with the parameters of table~\ref{FITKMN} 
\cite{HU01a}. 
Top left: Differential cross sections measured at 5.1$^\circ$ for $p+C$ (full squares) and 
for $p+Al$ (full circles) \cite{SU98} as a function of the particle momentum, compared with fit 
using \ref{FKALI} (thick solid for $Al$ and thick dashed for $C$), and using the KMN relation 
and parameters from \cite{KMN} (thin solid and thin dashed resp.). 
Top right: Same comparison for 14.6 GeV/c $p+Al$ invariant differential cross section versus 
mass transfer from \cite{AB93}: Fit with \ref{FKALI} (thick line) and KMN calculations \cite{KMN} 
(thin line).  
Bottom left: Same for 19.2 GeV/c  $p+p$ (full squares) and $p+Al$ (full circles) 
rapidity distributions from \cite{AL70}: fit with \ref{FKALI} (solid and dashed thick lines 
resp.) and KMN calculations \cite{KMN} (thin solid and dased lines resp.). 
Bottom right: 24 GeV/c $p+Al$ invariant differential cross sections at various angles (in msr
on the figure) from \cite{EI72} compared to the fits with \ref{FKALI}. For each measurement angle 
above the first (17~mrad), each next cross section has been multiplied by 10$^{-1,-2,...}$ for 
presentation purpose. KMN calculations are not shown on this figure for legibility.
Note that the same definitions of cross sections have been used as in the original references.
There is a clear target mass dependence of the differential cross section in the top left panel, 
while in the bottom left panel almost no such dependence is observed, because the observable 
displayed is a multiplicity, i.e., ratio of differential cross section to total reaction cross 
section.
\label{FITKMN} 
}
\end{center}
\end{figure*}

The invariant triple differential cross section is described by means of the formula used 
in \cite{KMN}:
\begin{equation}
(E \frac{d^3\sigma}{d^3p})_{inv} =
\sigma_R C_1 A^{b(p_t)} (1-x)^{C_2}\exp(-C_3x)\Phi(p_t) \label{FKALI} \\
\end{equation}
In this relation the kinematical variables $p_t$ and $x$ are the transverse momentum and 
the fractional energy of the particle respectively as defined in \cite{KMN} (relation 3.26 
in this reference), while $\sigma_R$ is the total $p+A$ reaction cross section. The function 
$\Phi(p_t)$ was modified as:
\begin{equation}
\Phi(p_t) = \exp(-C_4p_t^2) + C_5\frac{\exp(-C_6x_t)}{(p_t^2+\mu^2)^4}\exp(-\alpha\sqrt{s})
\label{PHI}
\end{equation}
With $b(p_t)= b_0\cdot p_t$ within the range of 4-momentum transfer considered here. 
Figure~\ref{FITKMN} shows some of the results obtained \cite{HU01a} by fitting the set 
of available data studied for this work with relation \ref{FKALI}. The energy dependent 
exponential factor in the second term of relation \ref{PHI} was introduced for it was found 
that this term contributes only at low incident energy. The values of the parameters 
obtained are given in table~\ref{TABKMN}. These values are significantly different from 
those given in the original work \cite{HU01a,HU02} on a smaller number of data. 
This work is currently being extended up to 400~GeV incident proton energy \cite{DU03}. 
\begin{table}
\begin{center}
\caption{\it Values of the parameters of relation \ref{FKALI} obtained in fitting the data 
of figure \ref{FITKMN}. %The $\chi^2$ per point obtained was around 1. 
\label{TABKMN}} 
\vspace{0.5cm}
\begin{tabular}{lllllllllll}
\hline
 Parameter    & $C_1$ & $C_2$ & $C_3$ & $C_4$ & $C_5$ & $C_6$ & $\mu$ & $b_0$ & $\alpha$ \\
 value        & 0.042 &  5.92 & 0.96  & 2.19  & 84.3  & 10.5  & 1.1   & 0.12  & 2.24  \\
\hline 
\end{tabular}
\end{center}
\end{table}
\subsection{$^4$He induced secondaries}

\subsubsection{Protons}

The inclusive $^4He+A\rightarrow p(n)+X$ proton (neutron) production cross section was obtained 
as described in section \ref{SIGPROT} above for the $p+A\rightarrow p+X$ cross section (using 
the total reaction cross section from \cite{JA78} for this system), renormalized to the 
available experimental multiplicities measured for this reaction \cite{AR87,BA93}.

\subsubsection{Antiprotons}

The inclusive antiproton production cross section was evaluated by means of the wounded 
nucleon model \cite{BI76,FA84} using the experimental values of the total reaction cross 
sections for the $^4He+A$ and $p+A$ systems, and of the \pb production multiplicity in 
nucleon-nucleon collisions \cite{HU02} available.

In this model, the particle production multiplicity $<n_{AB}>$ in the collision between 
ions $A$ and $B$ is related to the multiplicity nucleon-nucleon ${\cal(NN)}$ collision 
$<n_{\cal NN}>$ by the relation:  
\begin{equation}\label{eq:WNM}
<n_{AB}>~=~\frac{1}{2}\left(A \frac{\sigma_{pB}}{\sigma_{AB}} + 
                               B \frac{\sigma_{pA}}{\sigma_{AB}} \right) <n_{\cal NN}>
\end{equation}

with $\sigma_{ij}$ being the total reaction cross section between the $i$ and $j$ system.
Using this model, the \pb production multiplicity induced by the CR $He$ component on the 
nitrogen $N$ component of the atmosphere is found to be 
$<n_{HeN}>\approx 1.55<n_{pN}>\approx 2.5<n_{PP}>$.

\subsection{Total reaction cross sections}

$\bullet\;${\bf Protons:} The values of the proton total reaction cross section on nuclei 
used were obtained from the parametrization of \cite{LE83}, and checked on the carbon data 
from \cite{JA78}. \\
$\bullet\;${\bf $^4$He:} The $^4$He+A total reaction cross sections used were taken from 
\cite{JA78}.\\
$\bullet\;${\bf Antinucleons:} The \pb total reaction cross section was taken from 
\cite{CA79}, with the energy dependence from the data compilation of \cite{BA88}. The same 
production cross section and total reaction cross section have been assumed for \nb 
production as for \pb.

\section{RESULTS}\label{RESPBARS}
A sample of about 35 10$^6$ CRs have been simulated, of which 20\% were effectively 
propagated to the atmosphere (above GC), for detection altitudes going from sea level up to 
10$^4$ km altitude, including the BESS/CAPRICE balloon altitude ($\approx$38~km), the AMS 
orbit altitude ($\approx$370~km), and the recent BESS measurement terrestrial altitude 
(2770 m). The flux at sea level was calculated to investigate the possibility of ground level 
measurement of atmospheric antiprotons with existing devices \cite{HU01b}. This was achieved 
independently by BESS at mountain altitude and the results are discussed below. 
\subsection{Particle trajectories in the Earth magnetic field} %Lifetime of quasi trapped particles}
\begin{figure*}[htbp]          %fig 2: trajs
\begin{center}
%\vspace{2.0mm} 
\includegraphics[width=8.3cm]{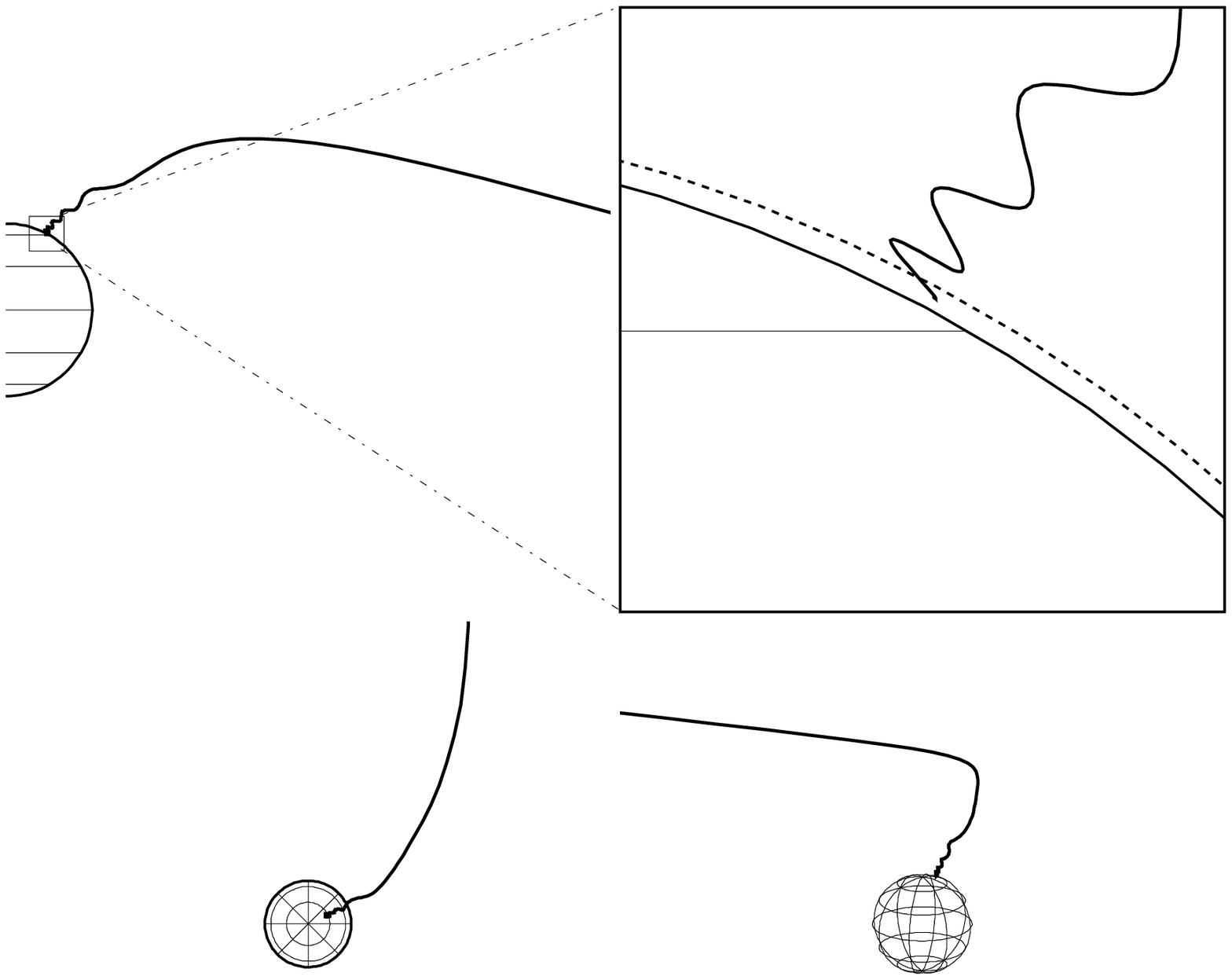} \includegraphics[width=8.3cm]{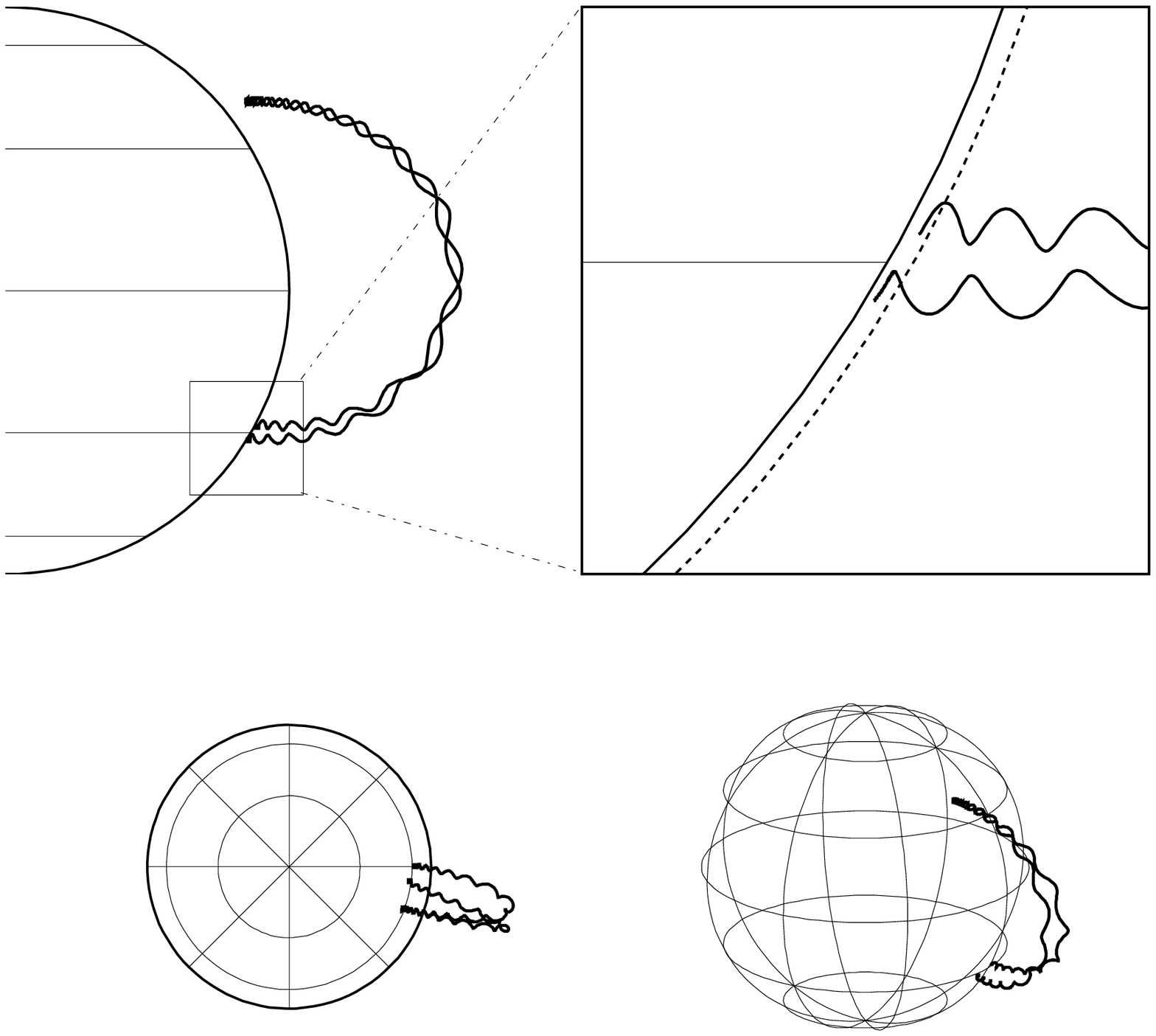} \\ 
\includegraphics[width=8.3cm]{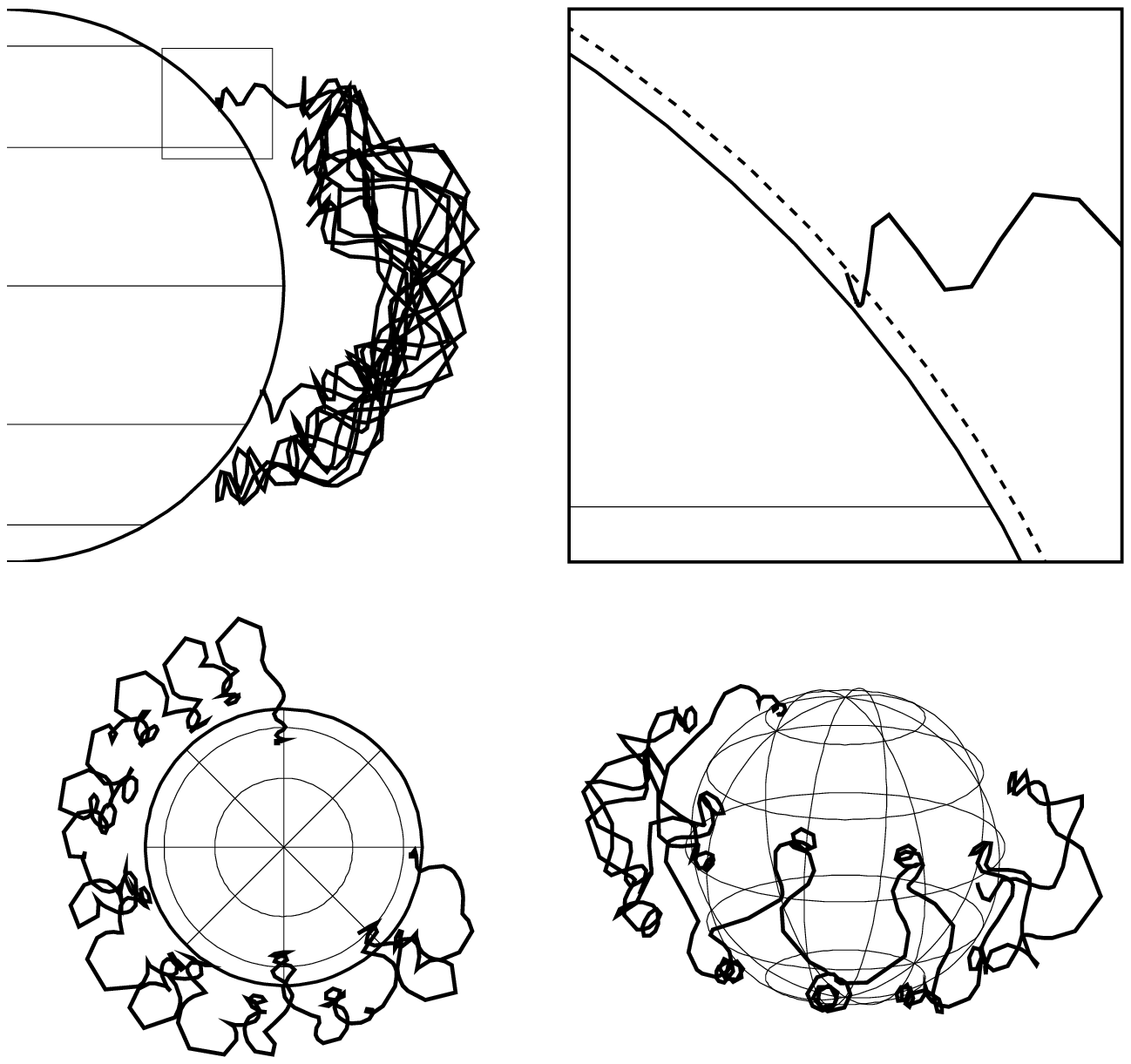} \includegraphics[width=8.3cm]{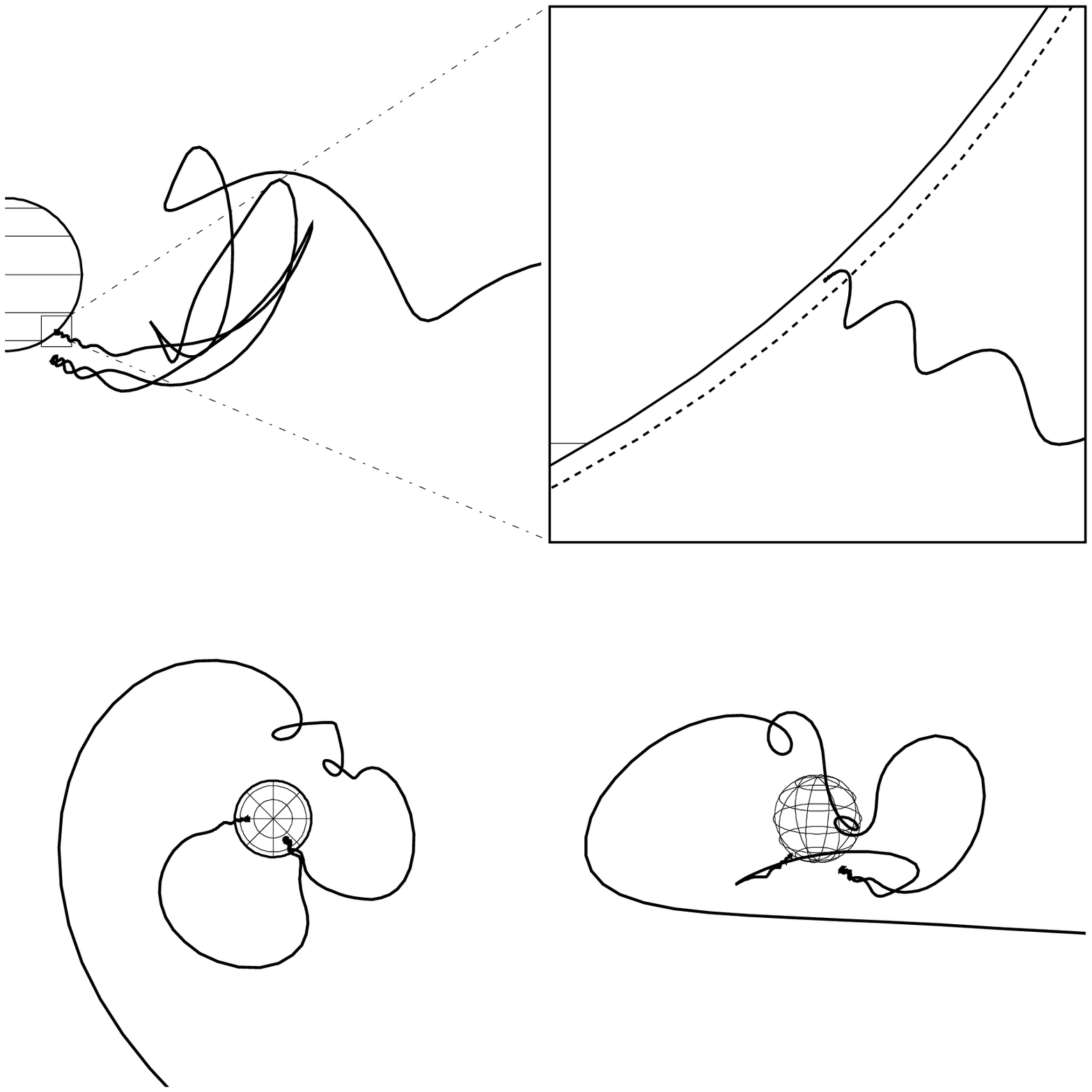} 
\caption{\it\small Examples of antiproton trajectories in the earth magnetic field. Details 
are discussed in the text. \label{TRAJS}}
\end{center}
\end{figure*}
The time of confinement of particles in the earth environment together with their 
particular trajectories, determine their status with respect to the three categories: 
trapped, semi-trapped, and non trapped (escape).
Trapped particles are spiralling back and forth around and along the magnetic field lines 
long enough to drift many times around the earth  (see for example \cite{SI62,ST55} and 
below). Trapped particles are practically not observed in the energy domain considered here. 
They are not dynamically forbiden however and a few trajectories with a few 10$^2$ bounces 
are observed, which corresponds to short-lived trapped particles. 
Quasi trapped particles are in similar kinematic conditions but accomplish only a limited 
number of bounces at mirror points before being absorbed or before escaping (see examples 
below). This concept appeared during the first years of radiation belts studies \cite{GA60} 
(see also the discussion in \cite{SI62}). 
Escape particles do not match the kinematic conditions for being trapped at their production 
point and escape in a very short time to the outer space. All intermediate situations 
between the stereotypes of quasi-trapped and escape trajectories are in fact observed in the 
simulation results (see example in bottom left fig~\ref{TRAJS}). 

Figure \ref{TRAJS} shows four examples of characteristic trajectories of antiprotons 
generated in this study. Each of the 4 panels gives a side view (projection on the meridian 
plane, top left), side view zoomed around the production point showing the spiralling 
trajectory of the particle (top right), top view (projection on the equatorial plane, bottom 
left), and 3-D representation (bottom right) of each of the selected trajectories. 

%The status of the trajectory: kinetic energy, rank, final state, lifetime and number of 
%bounces during lifetime, production altitude ?.

The top left event (1.52~GeV kinetic energy) is an escape particle produced close to the 
North pole. Top right is a semi trapped single bounce event (0.54~GeV) annihilating in the 
atmosphere close to its production point. Bottom left is a longer lifetime, multi bounces, 
semi trapped event (2.48~GeV), drifting around the earth about three quarters of a turn 
before annihilation in the atmosphere in the SAA region. Bottom right is intermediate 
between semi-trapped (since it displays at least one clear bounce) and escape event (0.54~GeV). 
It is a type of event for which the first adiabatic invariant (magnetic moment conservation) 
is not conserved because of a large variation of the magnetic field along the radius of 
gyration \cite{SI62}.
%Top left panel: 2.47 GeV escaping (final state 4) particle produced at high latitude.
%Top right: 1.25 GeV quasi trapped particle for which the (some ?) adiabatic invariant(s) is 
%not conserved (*** OK Laurent ?), thus escaping after a few bounces. Bottom left: 5 GeV 
%quasi trapped particle at high altitude, bouncing over a small range of low latitudes, and 
%drifting around the earth and finally annihilating in the atmosphere *** in the SAA ??*** 
%(final state 5). Bottom right: 1.92 GeV quasi trapped particle reaching close to 2 earth 
%radii in altitude and also disappearing in the atmosphere after ( n turns of drift around 
%the earth).
% \cite{GR77}  

\subsection{General features of the simulated data \label{DISCTR}} 

\begin{figure}[htb]          %fig 3: distribs générales
\begin{center}
%\vspace*{2.0mm} 
\hspace*{-0.5cm} 
\includegraphics[width=10cm]{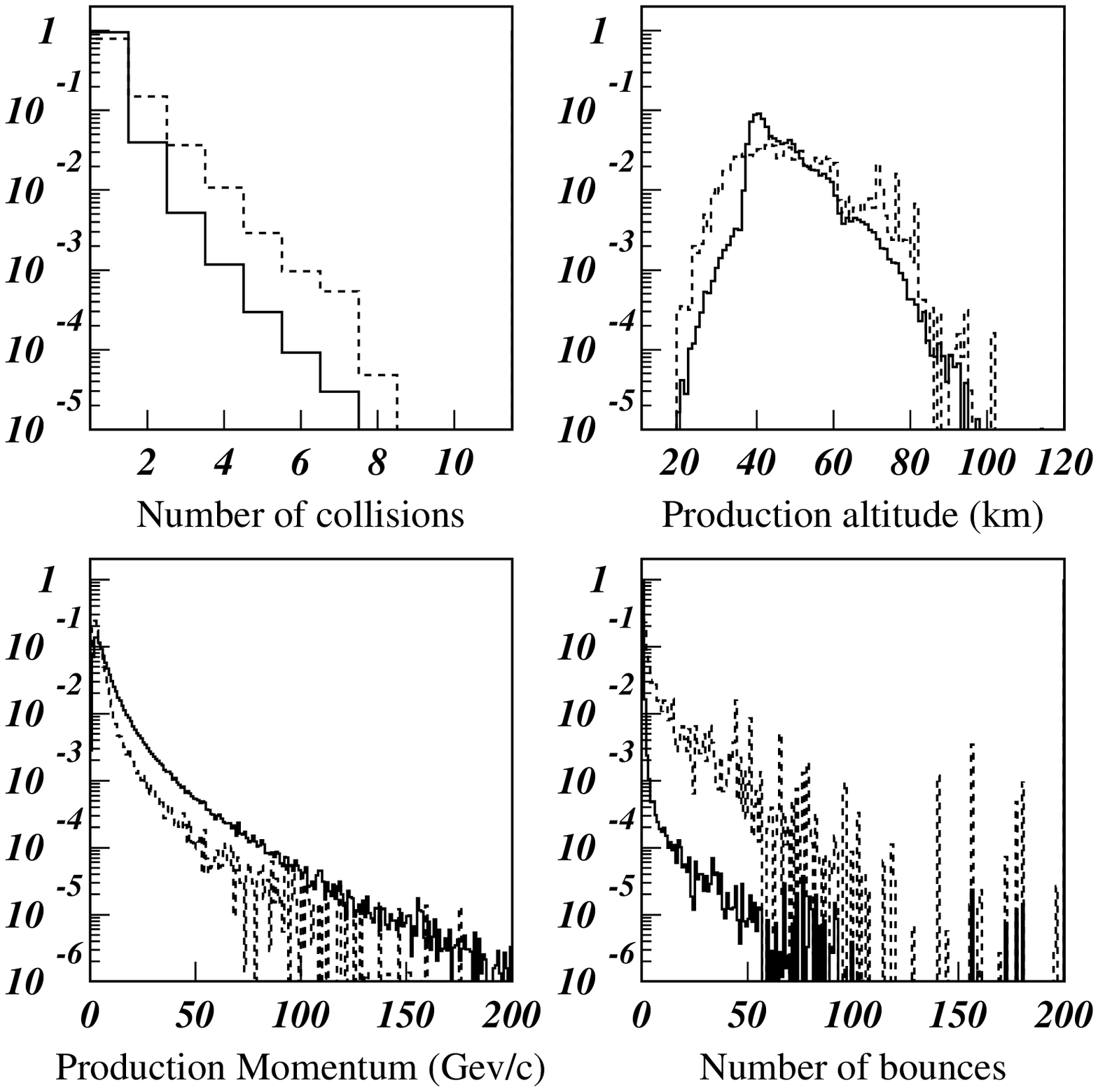}
\caption{\it\small General features of the simulated \pb sample at ballon (38~km) and 
satellite (380~km) altitude. Top left: Rank distribution (see text); Top right: Altitudes 
of production; Bottom left: Momentum distributions; Bottom right: Numbers of bounces effected 
by the particle between the mirror points. The spikes observed for high bounce multiplicity 
in this distribution correspond to quasi trapped particles trajectories crossing many times 
the detection altitude.
\label{DISTRIB} }
\end{center}
\end{figure}

Figure \ref{DISTRIB} shows a few basic distributions of physics observables relevant to the 
dynamics of the process for two detection altitudes: 38~km (solid line) and 380~km (dashed 
line) corresponding to balloon and satellite (AMS) altitudes, inside and outside the 
atmosphere, respectively. 
The rank of the collision producing the antiproton (top left panel on the figure) appears 
to extend from 1 up to about 10 for the simulated sample, showing that {\pbn}s are produced 
up to the tenth generation in the collisions cascade. The distributions are a little 
different for the two altitudes, with a significantly larger number of \pb occuring from 
first interaction at the lower altitude. 
The altitude distribution of the production point for the detection at 38~km (top right) 
shows a discontinuity at this altitude due to the incoming flux dominated by production from 
the upper layer of atmosphere. The mean production altitude is found around 46~km and 48~km 
for the lower and upper detection altitudes respectively. 
The particle momentum spectrum at the production point (bottom left) is found harder at the 
higher altitude. % why****.
The number of bounces at the mirror points for particle trajectories spiralling around 
the magnetic field lines are found as expected 
very different for the two altitudes of detection: At 38~km, only a small population is seen 
to reach a number of bounces larger than a few units (5-6). This flux is significant however 
and must correspond to trajectories lying mostly outside of the atmosphere. At 380~km, the 
observed flux of {\pb} trajectories with more than one bounce is larger by about 2 orders of 
magnitude than at 38~km, corresponding to the population of quasi-trapped particles as 
discussed previously in \cite{PAP1,BU01}.
\subsection{Antiproton flux at mountain altitude}
\begin{figure}[htbp]          %fig 4 *: Bess2770
\begin{center}
\includegraphics[width=8.3cm]{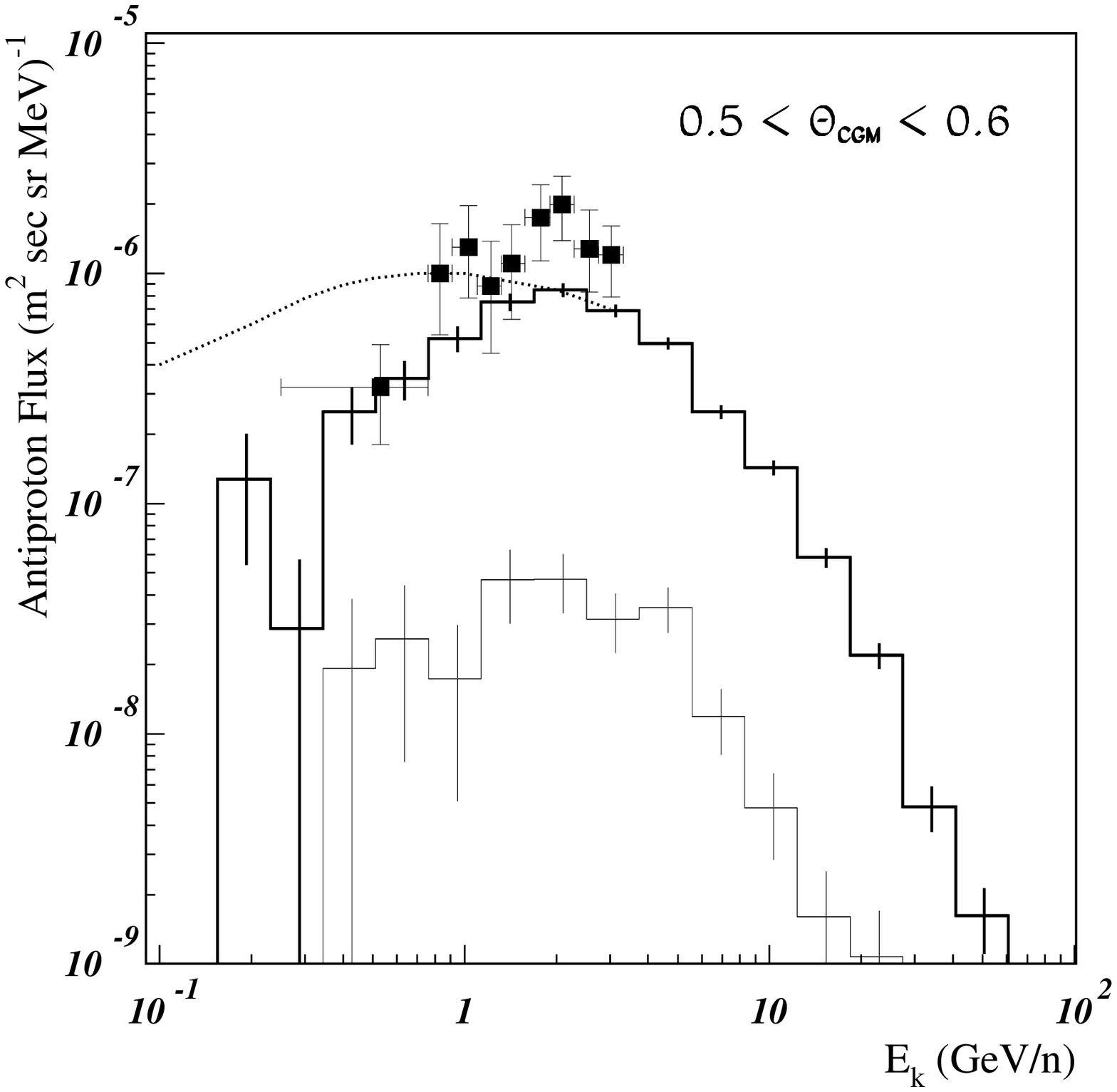}
\caption{\it\small 
Antiproton flux data at 2770 m measured by BESS \cite{FU01} (symbols) compared 
with simulation results (histograms). Thick histogram: full calculation; Thin histogram: 
CR $^4$He contribution; dotted line: transport equation calculation from \cite{ST97}.
\label{F2770}}
%\end{center}
%\end{figure}
%
%\begin{figure}[htbp]          %fig 5 *: Bess2770 exp zenith angle distrib
%\begin{center}
\includegraphics[width=8.5cm]{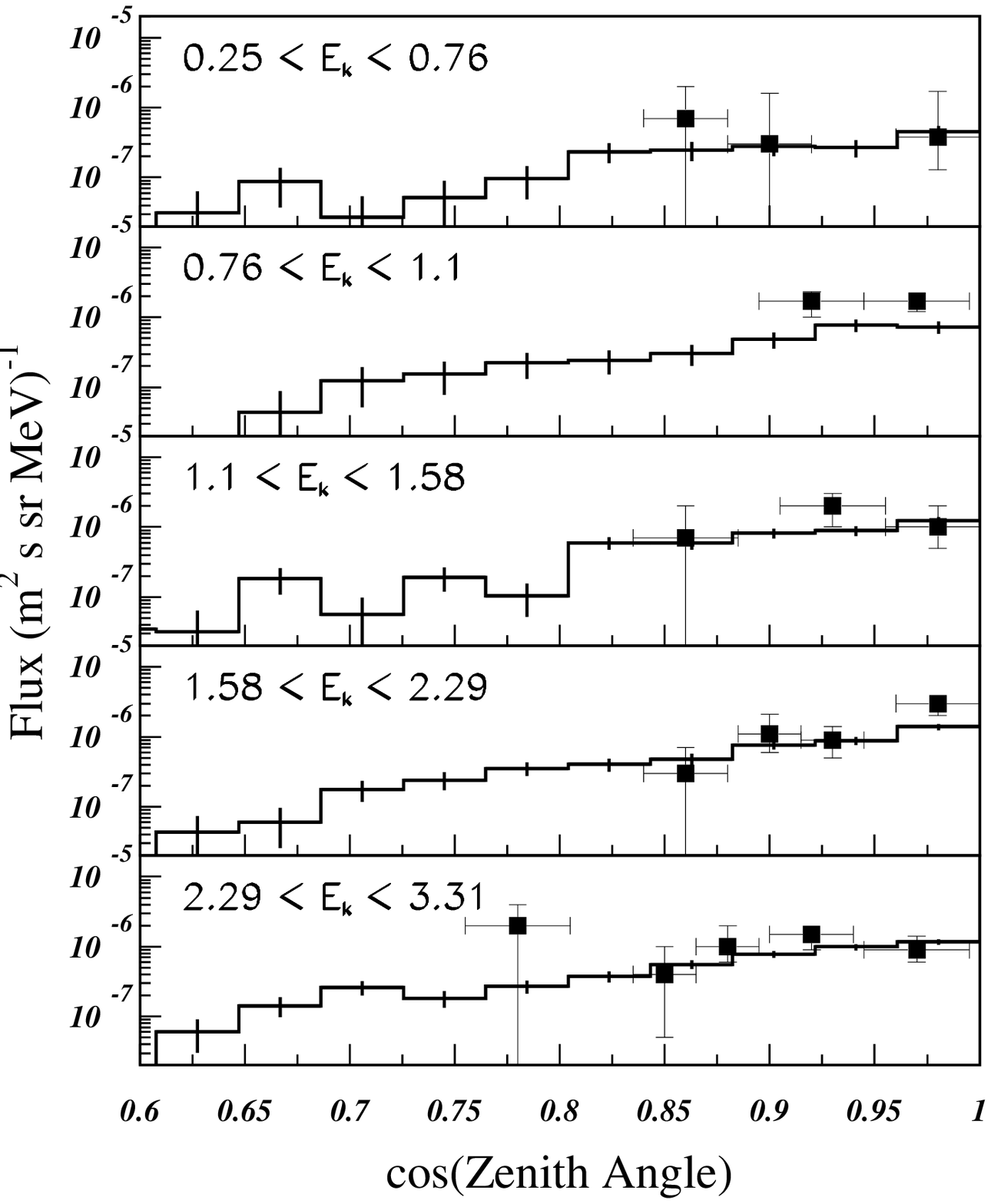}
\caption{\it\small Experimental zenith angle distribution of the antiproton flux at 2770 m 
in bins of kinetic energy compared with simulation results. 
\label{EZ2770}
}
\end{center}
\end{figure}

The recent measurements of the \pb flux at 2770~m of altitude by the BESS collaboration 
\cite{FU01} allows a sensitive test of the ability of the present simulation program to 
account for the observed flux since at this altitude the \pb production occurs on the average 
after a casade of 4 collisions on the average (see previous section and fig~\ref{DISTRIB}).
Another highly sensitive test of the overall calculation is provided in \cite{BA03} on 
the atmospheric proton flux.

Figure \ref{F2770} shows the \pb spectrum at 2770 m of altitude measured by BESS, compared 
to the simulation results. The latter has been run with the geometrical acceptance function 
of the BESS spectrometer given in \cite{FU01} (figures 4.37 to 4.39, see also \cite{AJ00}),
the overall acceptance angle being of the order of 25$^\circ$. 
The total CR $p+He$ and partial $He$ flux are shown on the same figure. The $H\!e$ flux 
contribution is seen to produce a small fraction of about 5\% of the full \pb flux. 
Although the total flux calculated somewhat underestimates the experimental values, the 
overall agreement is good, the calculated values being on the average within one standard 
deviation from the experimental values. This gives confidence in the results of the 
calculations obtained for the other altitudes investigated and reported below.

Figure \ref{EZ2770} compares the experimental zenith angle distributions of the \pb flux to 
the calculated values (histograms) for the same kinetic energy bins as measured by BESS 
\cite{FU01}. On this figure, the overall agreement between data and calculations again 
appears to be good for all energy bins.

Note that no upward particle have been produced at this altitude in the simulated sample, 
as it could be expected \cite{HU02}. 

\subsection{Balloon data}

In this section the atmospheric \pb flux at ballon altitude is investigated for comparison
with the atmospheric \pb corrections made to the raw flux data in the BESS and CAPRICE 
experiments.
 
\begin{figure*}[htbp]          %fig 6: Bess / caprice
\begin{center}
%\vspace*{2.0mm} 
\includegraphics[width=8.3cm]{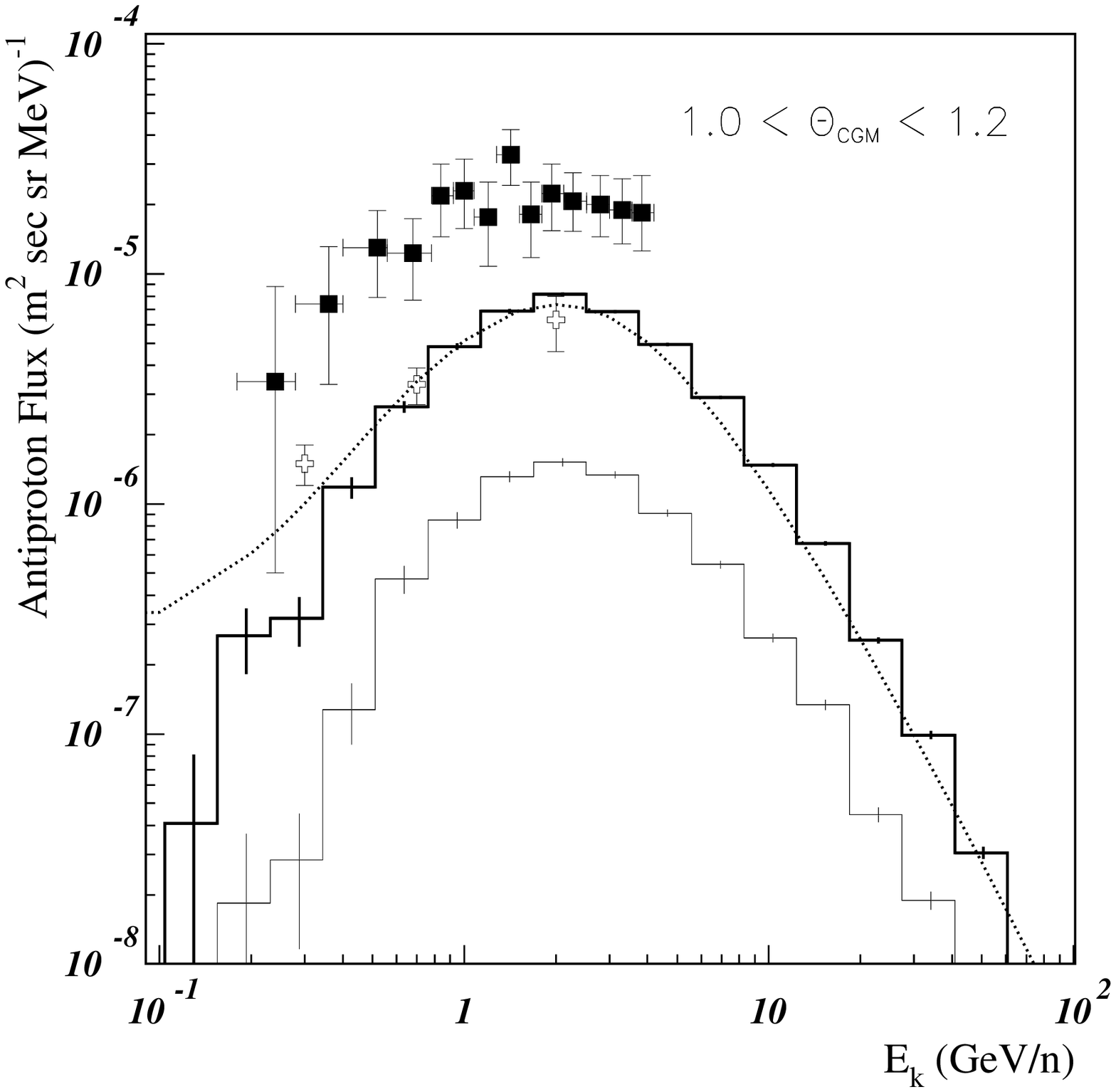}
\vspace*{2.0mm} 
\includegraphics[width=8.3cm]{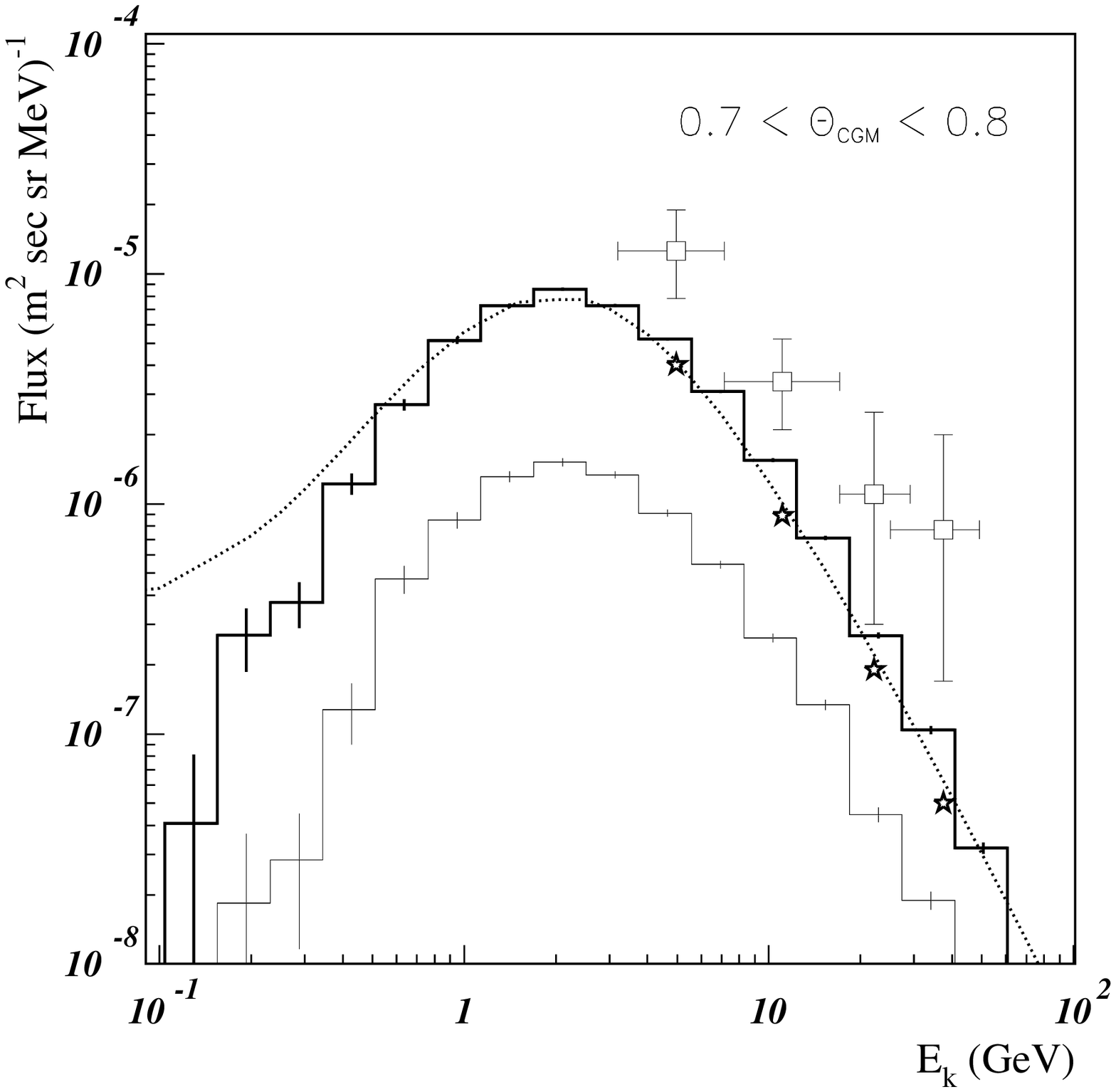} 
\caption{\it\small Left : BESS results: Galactic antiproton distributions deduced from the 
data (full squares), atmospheric antiproton flux from ref \cite{ST97} (curve), 
and corrections applied by the authors of \cite{BESS98} to correct the raw flux values for
the atmospheric contribution in the original work (open crosses), compared with the 
atmospheric flux obtained in the present work: all (CR proton + $^4$He) contributions (thick 
histogram); CR $^4$He contribution (thin histogram). Right : Same for the CAPRICE 
experiment data \cite{BO01} (data: open squares; corrections applied in \cite{BO01}: star 
symbols). In both panels $\Theta_{CGM}$ stands for the geomagnetic latitude of the 
measurements. 
\label{BESSCAP}}        %fig 7: ams
\vspace*{2.0mm} 
\includegraphics[width=8cm]{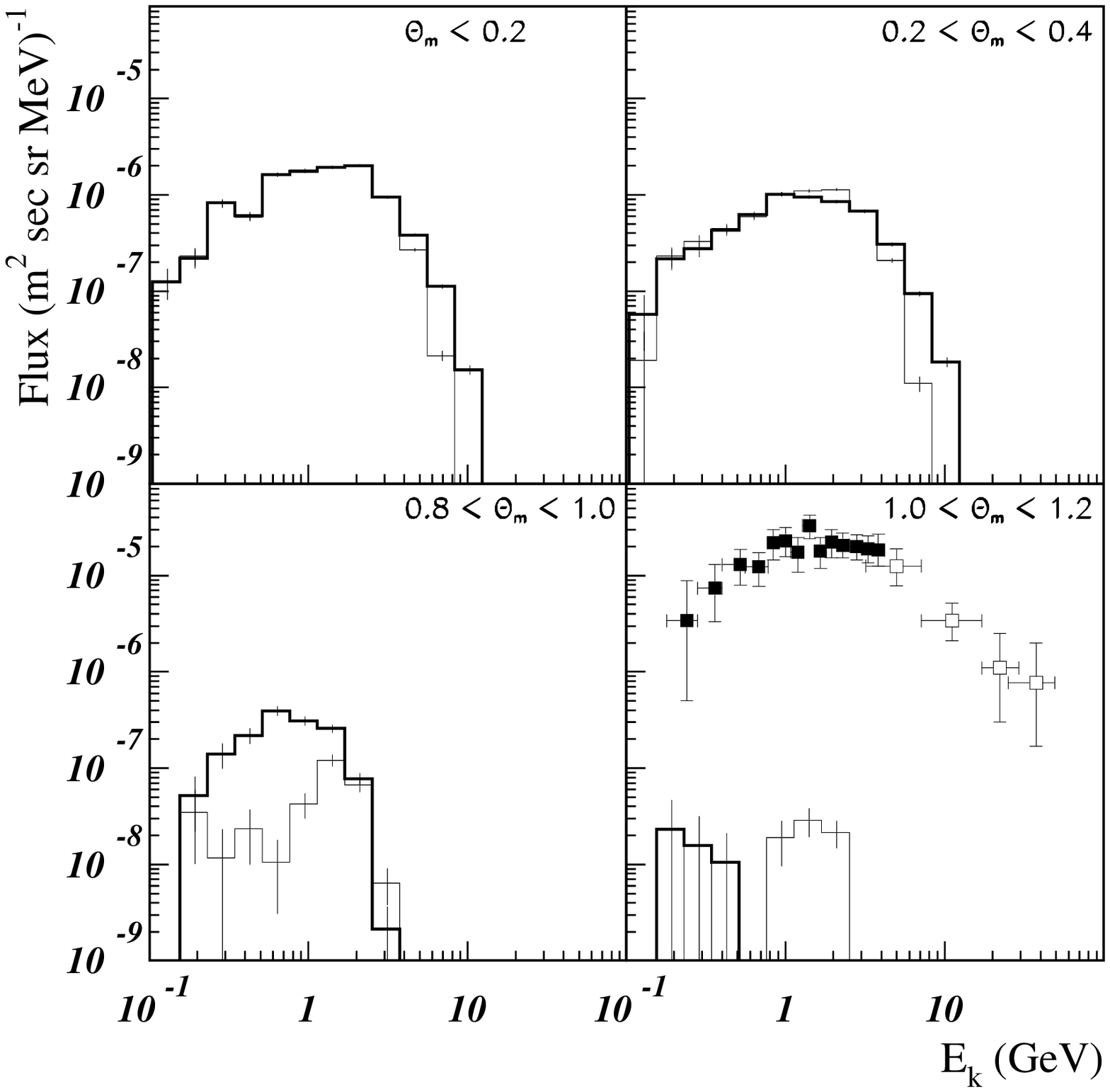} 
\caption{\it\small Atmospheric antiproton spectra expected from the present work for the 
AMS experiment on the ISS at 380~km of altitude for 4 bins of latitudes, 
compared with the AMS, BESS and CAPRICE data \cite{AMS1T,BESS98,BO97} in the 
polar region (bottom right). 
Full line: Downward flux; Dashed line: Upward flux. The increasing flux with decreasing 
latitude is due to the larger (quasi-)trapped particle population closer to the equator 
(see text).
\label{AMS1}}
\end{center}
\end{figure*}

Figure~\ref{BESSCAP} shows the values of the galactic \pb flux obtained from the BESS and 
CAPRICE measurements respectively. These values have been obtained from the measured raw 
flux by subtraction of the atmospheric \pb flux evaluated using an average of theoretical 
calculations for the BESS experiment \cite{BESS98}, and using the calculations of ref 
\cite{ST97} for CAPRICE. On the figure, the atmospheric flux calculated in \cite{ST97} is 
compared with the results from the present work (see also \cite{PF96}).
For the two sets of data, it appears that the the present calculations are in fairly good 
agreement with the atmospheric antiproton flux obtained from transport equation calculations 
and used to correct the raw flux measured. There is a slight trend however for the 
simulation results to be larger than those obtained from the differential equation approach 
by about 20\% over the range 10-30 GeV.  
At low \pb energies the opposite trend is observed and the simulation results are found 
significantly below the values obtained from the differential equation. One might say that 
the simulation results should be taken with care below 1~GeV because of the lack of 
experimental cross sections for low energy \pb production, and thus of the large 
corresponding uncertainties on the results of the simulation over this range, 
%(This must affect the Stephens calculs as well ???)
however, it must be noted that the calculated cross sections for low \pb momentum should 
be in principle reliable for the following reason. 
The \pb distribution is naturally symmetric in the rapidity space. The fitting function has 
the same property since it depends only on variables matching this symmetry. Therefore a 
good fit to a set of experimental cross sections for \pb rapidities above the center of 
mass rapidity $Y_{cm}$, %i.e., particle momenta above the center of mass momentum, 
automatically ensures the right behaviour of the calculated values for rapidities below 
$Y_{cm}$, i.e., for particle momenta in the laboratory frame, because of the symmetry law.

To conclude this section, the atmospheric \pb flux calculated in this work confirm the 
corrections of the raw flux values measured in the BESS and CAPRICE experiments.
This result updates and corrects a previous preliminary conclusion on the issue \cite{HU01b}
recently quoted in \cite{MO03}. 

The contributions of A$>$4 CR components were not included in the calculations, 
neither were those from the non-annihilating inelastic contributions in the \pb propagation 
through the atmosphere. These contributions are small however and not likely to change the 
results by more than a few percents \cite{LI03}.

\subsection{AMS altitude}

Future satellite experiments in preparation, plan to measure the \pb flux. A reliable
knowledge of the atmospheric \pb flux at satellite altitudes is therefore highly desirable
for these experiments to step on explored grounds. The \pb flux calculated for the altitude
of the AMS orbit are presented in this section.

Figure~\ref{AMS1} shows the expected downwards (Secondaries and reentrant Albedo, dashed 
histogram) and upwards (Splash Albedo particles, solid histogram) flux of atmospheric 
antiprotons at the altitude of AMS for two regions of lower geomagnetic latitude: equatorial 
($0<|\theta_M|<0.2$~rad), intermediate ($0.2<|\theta_M|<0.4$~rad), and subpolar 
($0.8 <|\theta_M|<0.9$~rad). As expected, the flux is predicted larger for the lower 
latitudes than it is around the poles, because of the existence of quasi-trapped 
\pb components at the low and intermediate latitudes. Note that the simulated flux is 
surprisingly predicted larger downwards than upwards (bottom left panel). This is in fact 
an effect of the spectrometer acceptance (taken to be 30$^\circ$ with respect to zenith), 
the mean angle for upward particle trajectories being 2 radians. The overall upward flux 
is larger than downward by a factor of about 2.5 \cite{HU01b,HU02}. 
This shows that the future satellite measurements of antiproton flux at low latitudes will 
have to be corrected from the atmospheric contributions and will probably suffer more 
uncertainties than previously thought.

The lower right panel compares the \pb data at TOA reported by AMS \cite{AMS1T}, BESS 
\cite{BESS98}, and CAPRICE \cite{BO01}, to the flux calculated in the polar region where 
the AMS data have been measured. The calculated (downward) atmospheric \pb component is at 
the percent level of the measured flux in the low energy \pb range, and can be considered as 
negligible at all energies.
\begin{figure*}[htbp]          %fig 8 : pbars nbars
\begin{center}
%\vspace*{2.0mm} 
\includegraphics[width=8cm]{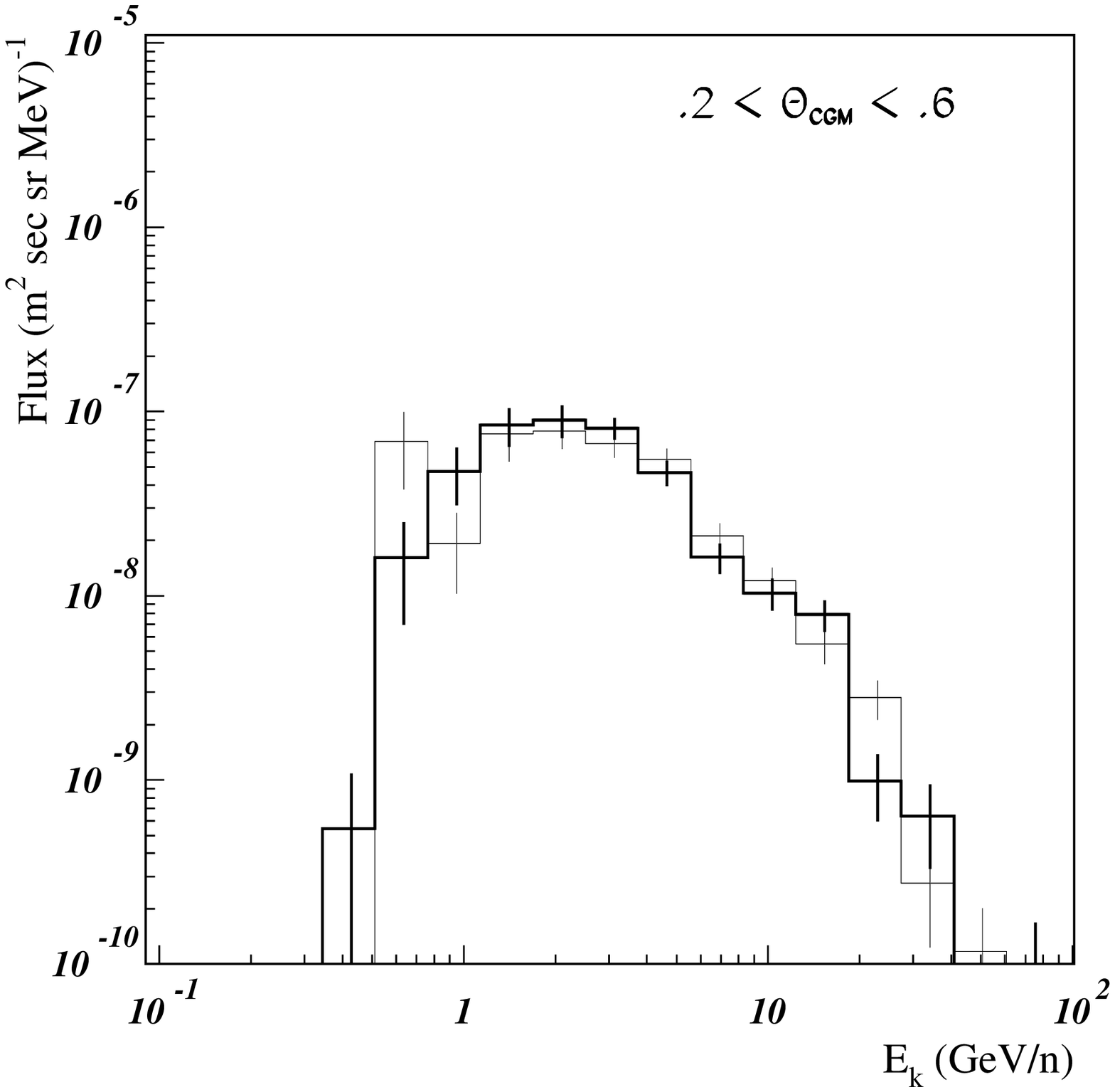} 
\includegraphics[width=8cm]{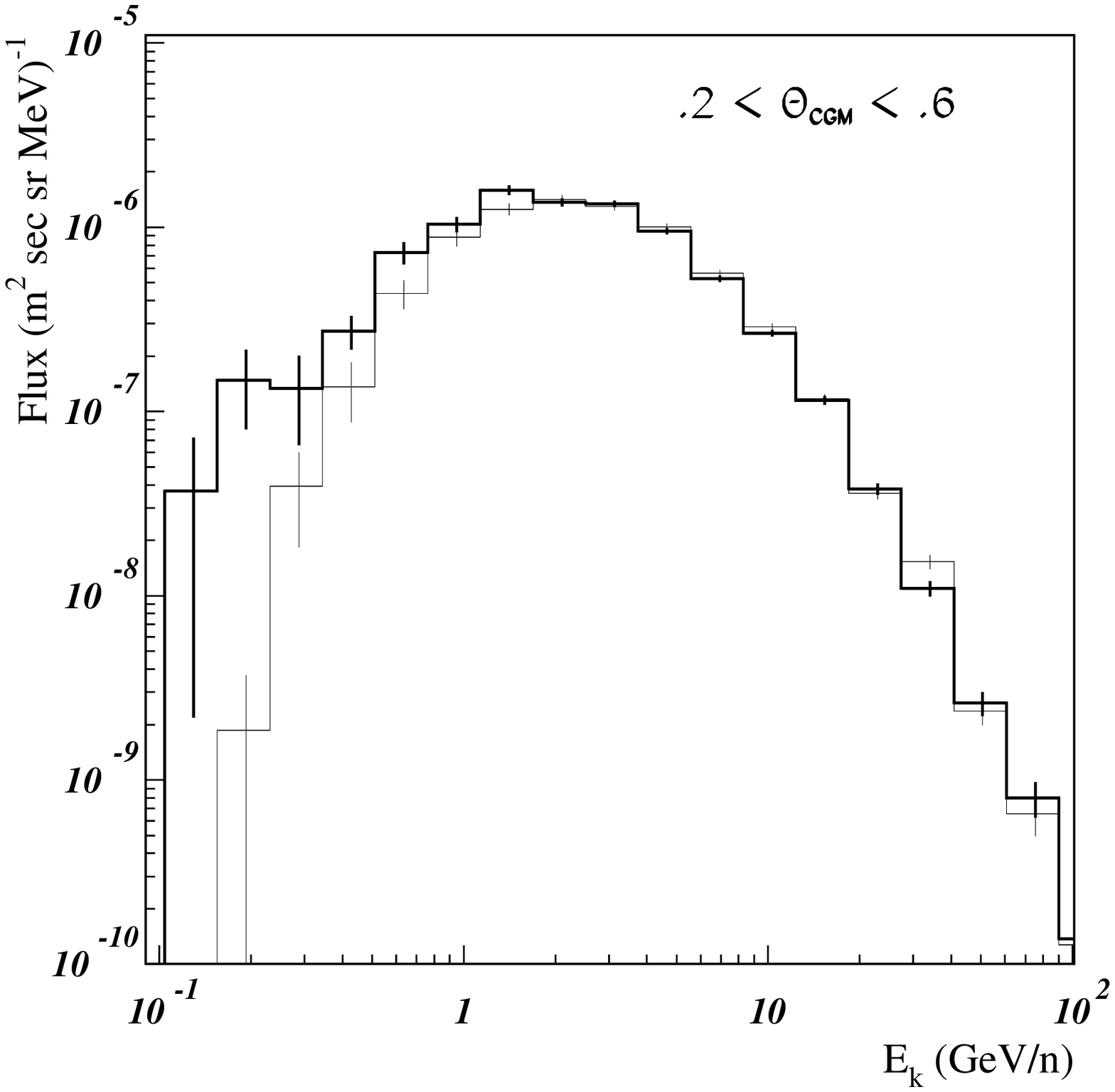} 
\caption{\it\small Simulation results for the antiprotons (thick line) and antineutrons 
(thin line) flux at sea level (left) and at 4000~meters of altitude (right). \label{SEA4}}
\end{center}
\end{figure*}
\subsection{\pb and \nb flux at terrestrial altitudes}

The \pb and \nb flux have been calculated also at sea level in order to provide a realistic
order of magnitude of these flux for general purpose and for ground testing of embarked 
experiments.

$\bullet\;$Antiprotons : 

The flux of atmospheric antiprotons at sea level has been calculated with the same 
simulation program. Figure~\ref{SEA4} shows the distributions obtained at sea level (left)
and at 4000~m. The energy integrated flux is of the order of 0.4 10$^{-3}\,$~\pb $s^{-1}
m^{-2}sr^{-1}$ at all latitudes (see fig \ref{FLXALT} below). At 4000~m (right panel on the 
figure), the flux raises to about 7 10$^{-3}\,$~\pb $s^{-1}m^{-2}sr^{-1}$. These values are 
small but large enough for this flux to be measured by currently existing large acceptance 
detectors (BESS, CAPRICE), or in a near future by new detectors under construction like AMS 
and PAMELA. 

$\bullet\;$Antineutrons : 

Atmospheric secondary antineutrons may also be of interest in ground or balloon measurements 
\cite{ME01}. Figure~\ref{SEA4} shows the kinetic energy spectrum of the expected \nb flux at 
sea level and at 4000~m (this latter altitude being that the Cerro-La-Negra observatory in 
Mexico where some experimental measurements of the antineutron flux are being considered 
\cite{CERRO}).
\begin{figure}[htbp]          %fig 9 ** : Flux vs altitude
\begin{center}
%\vspace*{2.0mm} 
%\includegraphics[width=8.3cm]{flux_alt.eps} 
%\includegraphics[width=11cm]{intflux.eps} 
\includegraphics[width=8.5cm]{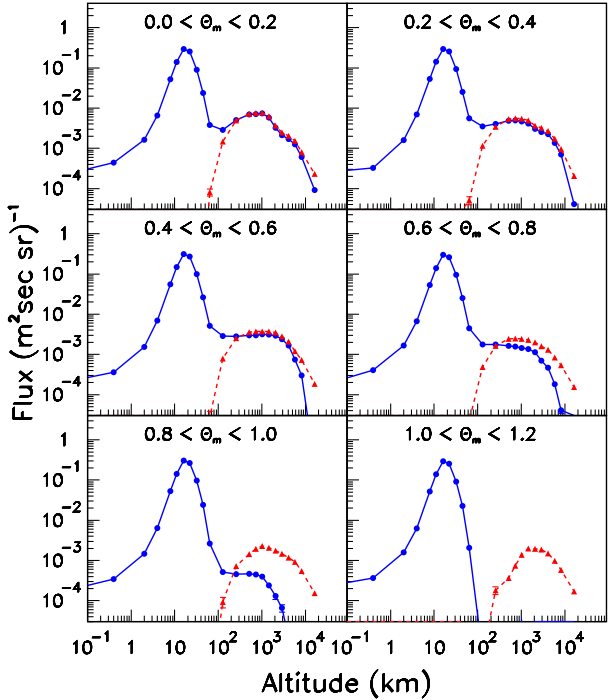} 
\caption{\it\small Integrated antiproton flux versus detection altitude in bins of latitudes 
between equator and poles, for downward (solid line) and upward (dashed line) \pb flux. 
\label{FLXALT}}
\end{center}
\end{figure}

\subsection{Flux dependence on the altitude}\label{HIGHALT}
The \pb flux has been calculated up to altitudes of 2 10$^4$~km with the aim of 
investigating the general features of the dynamics and kinematics of the particles in the 
more remote Earth environment than considered in the previous sections. 

Figure \ref{FLXALT} shows the altitude dependence of the energy integrated upward and 
downward \pb flux in bins of latitude, assuming a geometrical acceptance of 30$^\circ$ for 
the detector. The calculated distributions display two main features: \\
1) In the atmospheric range of altitudes, a large peak of incoming flux centered around 
20~km and corresponding to atmospheric secondaries, dominates the distribution 
($\lesssim$~50km, i.e., $\lesssim$TOA), with basically no associated outgoing flux (inside 
the quoted acceptance angle).\\
2) Above the atmosphere, surprisingly, the calculated upward and downward flux are 
found close to each other up to fairly high altitudes, namely $\approx$10$^4$~km, for the 
low and intermediate latitudes ($|\theta_M|\lesssim 0.7$~rad). This shows that a 
population of quasi-trapped, particles should be observed in this region of space, i.e., 
up to around five to ten thousands kilometers. This is confirmed by the lifetime of the 
particle between their production and their absorption, and by the number of bounces of 
the particles between the mirror points of their trajectories, which extend up to 100 
seconds and several 10$^2$ bounces (see Fig.~\ref{DISTRIB}), respectively for the simulation 
sample produced. 
At higher latitudes with ($|\theta_M|\gtrsim 0.7$~rad), the incoming flux progressively 
disappears, and the outgoing flux then corresponds to escape particles.

From these calculations it can be concluded that there should exist a significant flux of 
quasi-trapped particles extending approximately over a decade of altitudes, from about 
50~km (TOA) up to $\approx$10$^4$~km, depending on the particle energy and latitude. This 
flux has been observed already by the AMS experiment in the lower part of the altitude 
range (380~km) \cite{PAP1}. Note that the issue was discussed long ago in a pioneering 
paper about the electron flux \cite{GR77}. See also \cite{BA03} (companion paper) for a 
complementary discussion. 
The energy spectrum of this flux extends up to around 10~GeV, which is about the upper 
momentum limit (8.5~ GeV) for which particles can match the simple geometrical condition 
that the gyration radius is smaller than the mean trajectory radius to the upper 
atmosphere (for equatorial latitude trajectories, at the limit of large, close to $\pi$/2, 
pitch angles).

%a value at which the gyration 
%radius of {\pbn}s is of the order of 1000~km. The relative variation of the magnetic field 
%around the equator is about ***\%, for which the first adiabatic invariant is thus poorly 
%conserved.
%a natural limit since the gyration radius of charge one particles is of the order of 100~km 
%per GeV  

Future embarked experiments should take these features into account, even though, in 
principle, the accurate knowledge of the kinematical conditions of the particle at the 
detection point allows to know whether it is or not of atmospheric origin. 

\section{Summary and conclusion}\label{CONC}
In summary, the secondary antiproton flux produced by the Cosmic Ray proton and Helium flux 
on the atmosphere has been calculated by Monte-Carlo simulation. 
The flux calculated for the altitude of 2770~m is in fair agreement with the recent BESS 
measurements. At sea level, it is small but measurable and could provide a natural facility 
for testing the identification capability of existing experiments or of future devices.
For balloon altitudes, the calculated flux has been found in agreement with the values 
calculated in previous works. At satellite altitudes (380~km) it appears to be negligible 
compared to the CR flux for polar latitudes, and of the same order of magnitude as for the 
high balloon altitudes for equatorial and intermediate latitudes (below the geomagnetic 
cutoff), indicating that it will have to be taken into account in future measurements of the 
galactic antiproton flux at similar altitudes.  

{\sl Acknowledgements:}
The authors are indebted to V. Mikhailov for helpful discussions on the issue and for 
pointing ref \cite{GR77} to them. They are also grateful to M.~Nozaki, M.~Fujikawa, 
and the BESS collaboration, for providing their measurements of \pb flux at 
2770~m prior to publication.

\end{document}